# Ctrax extensions for tracking in difficult lighting conditions

Ulrich Stern[1,3] and Chung-Hui Yang[2,3]

[1] Independent researcher, ulrich.stern@gmail.com
[2] Dept. of Neurobiology, Duke University, yang@neuro.duke.edu
[3] Corresponding author

**Abstract**
The fly tracking software Ctrax by Branson *et al.* is popular for positional tracking of animals both within and beyond the fly community.  Ctrax was not designed to handle tracking in difficult lighting conditions with strong shadows or recurring "on"/"off" changes in lighting – a condition that will likely become increasingly common due to the advent of red-shifted channelrhodopsin.  We describe Ctrax extensions we developed that address this problem.  The extensions enabled good tracking accuracy in three types of difficult lighting conditions in our lab.  Our technique handling shadows relies on "single animal tracking"; the other techniques should be widely applicable.

The fly tracking software Ctrax by Branson *et al.*[1] is popular for positional tracking of animals both within and beyond the fly community.  Ctrax offers a multitude of options that allow tailoring it to the particular tracking task at hand.  For tracking in difficult lighting conditions with strong shadows or recurring "on"/"off" changes in lighting – a condition that will likely become increasingly common due to the advent of red-shifted channelrhodopsin, however, Ctrax can make a prohibitive number of tracking errors.  The Ctrax extensions described here address difficult lighting conditions.  We study how female *Drosophila* decide between two egg-laying sites[2] and typically track the flies for 8 hours; we needed the Ctrax extensions in two types of experiments: when illuminating one of the sites with UV light either constantly ("UV on") or periodically ("UV on/off") (**Fig. 1a-b**) and when illuminating the entire egg-laying chamber with strong red light recurrently to optogenetically activate neurons (with ReaChR[3, 4]) (**Fig. 1c-d**).

Ctrax detects flies based on the difference between the current frame and the "background" without flies, and shadows (and reflections) of flies can have differences comparable to the flies' differences, causing false positives (**Fig. 1e-h**).  For our "UV on" experiments with strong shadows, unmodified Ctrax typically detected hundreds of flies over the course of 8h instead of the two flies (one per chamber) we record per video (**Fig. 1i**).  We extended Ctrax with a shadow detector, which, for each frame, discards all flies but the ones closest to the center of each chamber, taking advantage of the chamber geometry and that we have one fly per chamber (**Fig. 1g-h**, **Supplementary Fig. 1**).  If the background changes gradually over time, the shadow detector – and Ctrax in general – can be prone to mistakes, which we addressed by recalculating the background typically every 30 or 60 minutes (**Supplementary Fig. 2a**).  To assess tracking accuracy, we examined the flies' trajectories and implemented a detector for suspicious jumps (**Supplementary Fig. 3a-b**).  Shadow detector and background recalculation enabled tracking "UV on" for 8h with good accuracy (**Fig. 1i**, **Supplementary Fig. 3e**).

Unmodified Ctrax cannot handle tracking with recurring changes between two different lighting states ("on"/"off") since it uses a single background. We hence extended Ctrax with a simple on/off detector. The detector randomly picks 100 frames, calculates the mean (average brightness) for each, and classifies the means into two clusters ("on"/"off") via k-means. If the cluster centroids differ by more than 3% in brightness, the detector assumes there are two lighting states, a separate background is calculated for each state (**Supplementary Fig. 2b**), and the right background is chosen for each frame during tracking. The detector also extends the Ctrax trajectory output file with the information about the timing of the lighting state changes it learned from the video. Our suspicious jump detector regularly detected tracking errors coinciding with lighting state changes, so we extended it to automatically fix such errors (**Supplementary Fig. 3c-d**). Combining our extensions enabled tracking with good accuracy in both our "UV on/off" (**Fig. 1j**) and "strong red light on/off" experiments.

Our Ctrax extensions (based on Ctrax 0.3.1) are available as project yanglab-ctrax on Google Code (https://code.google.com/p/yanglab-ctrax/). The extensions are relatively simple and worked well for our experiments. Our shadow detector relies on having a single animal "per chamber." Recent more advanced techniques[5,6] *may* enable reliable shadow detection in multiple animal tracking at additional implementation cost. Our other extensions – background recalculation, on/off detector, and auto-fixing jump detector – should be widely applicable.

## Acknowledgments

We thank Edward Zhu and Bin Gou for performing the experiments whose videos were used for this report and E.Z. for his comments on a draft of this paper. C.-H.Y. was supported by the National Institutes of Health under award number R01GM100027.

1. Branson, K., Robie, A.A., Bender, J., Perona, P. & Dickinson, M.H. High-throughput ethomics in large groups of Drosophila. *Nature methods* **6**, 451-457 (2009).
2. Yang, C.H., Belawat, P., Hafen, E., Jan, L.Y. & Jan, Y.N. Drosophila egg-laying site selection as a system to study simple decision-making processes. *Science* **319**, 1679-1683 (2008).
3. Inagaki, H.K. et al. Optogenetic control of Drosophila using a red-shifted channelrhodopsin reveals experience-dependent influences on courtship. *Nature methods* **11**, 325-332 (2014).
4. Lin, J.Y., Knutsen, P.M., Muller, A., Kleinfeld, D. & Tsien, R.Y. ReaChR: a red-shifted variant of channelrhodopsin enables deep transcranial optogenetic excitation. *Nature neuroscience* **16**, 1499-1508 (2013).
5. Krizhevsky, A., Sutskever, I. & Hinton, G.E. in Advances in neural information processing systems 1097-1105 (2012).
6. Perez-Escudero, A., Vicente-Page, J., Hinz, R.C., Arganda, S. & de Polavieja, G.G. idTracker: tracking individuals in a group by automatic identification of unmarked animals. *Nature methods* **11**, 743-748 (2014).

**Figure 1** Tracking in difficult lighting conditions and performance of the Ctrax extensions.

**(a)** Sample frame from our "UV on" experiments, showing two chambers. For the left chamber, the top edge of the chamber sidewall is outlined in yellow, and the two egg-laying sites at the bottom of the chamber are outlined in white. The blue arrow points to a UV LED (below the chamber in our setup). Note that there is one fly per chamber (white arrows). A red "light pad" provides additional lighting – that is invisible to *Drosophila* – for tracking.

**(b)** Sample frame from our "UV on/off" experiments at a time when the UV is off. With UV on, the frame would look similar to (**a**). The small dark spots on the egg-laying sites are eggs (arrows).

**(c-d)** Sample frames from our experiments using strong red light to optogenetically activate neurons, with arrows pointing to flies. **(c)** Two long chambers (one outlined in yellow) illuminated with strong red light; there is no red visible in the image since – to reduce light intensity for the camera – a filter that lets only blue light (400-500nm) pass (LEE Filters, 713 J.Winter Blue) was placed in front of the camera. **(d)** Same chambers as in (**c**) with red light turned off.

**(e-h)** Sample frame with strong shadows that led to false positives, with white and green arrows pointing to flies. **(e)** Frame in grayscale, which Ctrax uses for tracking. **(f)** Difference between frame (**e**) and background ("frame without flies"), with darkness proportional to the absolute value of the difference. The shadows (blue arrows) of the right fly have a larger difference than the fly itself. (We used Ctrax's "Background Brightness" normalization, which performed best for our chambers.) **(g)** The same difference as in (**f**) is now shown in green and superimposed onto the background. **(h)** Flies detected by Ctrax shown as ellipses. For each chamber, our shadow detector picks only one ellipse (fly) – the one closest to the center (yellow arrow) of the chamber, which eliminates all false positives in this frame.

**(i)** Results of tracking using unmodified Ctrax and Ctrax with extensions on four "UV on" sample videos (8h each). Ctrax with extensions correctly detected just two flies for each video, while unmodified Ctrax detected hundreds of flies. We manually examined the 12 jumps our suspicious jump detector reported for the four videos (**Supplementary Fig. 3b**), and tracking was correct in all cases (i.e., the flies did jump in these cases). Note that neither table (**i**) nor (**j**) lists "minor" tracking errors that were below the detection threshold of the suspicious jump detector (**Supplementary Fig. 3e**).

**(j)** Results of tracking using Ctrax with extensions on three "UV on/off" sample videos (8h each). (We did not run unmodified Ctrax since it was not designed to handle "on"/"off" changes.) "1 min on/off" was 1 min "on," 1 min "off," 1 min "on," etc. The correct number of flies was detected for each video. We manually examined all jumps our suspicious jump detector either automatically fixed (**Supplementary Fig. 3c-d**) or just reported. Of the 44 jumps it automatically fixed for the three videos, it made one error but was correct in the remaining 43 cases. So the "auto fix" feature strongly reduced tracking errors here. For the 9 jumps the detector just reported, tracking was correct.

**Supplementary Figure 1** Template matching – how the shadow detector knows where the centers of the chambers are.

**(a-b)** Template matching is used to determine the exact position of the chambers in a particular video, which, in turn, yields the positions of the centers of the chambers needed by the shadow detector (**Fig. 1h**). We use a total of 18 cameras (Microsoft LifeCams) for recording *Drosophila*, which made the chamber position vary by video. (In situations with little variation in chamber position, hard-coding could be used instead of template matching.) **(a)** Template image used to detect chamber position in video. **(b)** Example of a match with the white rectangle indicating the edge of the template in the position that best matches template and background. We used the background (without flies) so that flies (and possibly their shadows) would not interfere with the match. Our template matcher uses multiple image transformations (primarily different Canny edge detections), enabling it to reliably match a template from good lighting conditions against backgrounds from a wide range of lighting conditions. (We use the same template matching code also in scripts to analyze behavior for experiments in both good and difficult lighting conditions.)

**Supplementary Figure 2** Background recalculation for "UV on" and for "UV on/off."

**(a)** Background recalculation for an 8h "UV on" video. The background calculated over the 1st hour is shown at the top of the panel (bg1). Differences between bg1 and the backgrounds calculated over the 2nd and 8th hours ($\Delta$bg2,bg1 and $\Delta$bg8,bg1) are shown below bg1. Some of the background changes here (arrows) are caused by changes in grape juice level; to increase egg-laying, we typically provided the flies with grape juice in a small well in the center of each chamber. If only a single background is calculated over the whole 8-hour video, the changing juice level can lead to false positives, which the shadow detector typically picks over the fly since the well is in the center of the chamber. This mistake, in turn, causes flies to appear to be jumping between their actual positions and the center (**Supplementary Fig. 3a-b**).

**(b)** Backgrounds calculated by the on/off detector for an 8h "UV on/off" video. Separate backgrounds for "on" and "off" were calculated over each hour, with the backgrounds for the 1st hour shown at the top of the panel (off1, on1). Differences between off1 and the backgrounds for "off" calculated over the 2nd and 8th hours ($\Delta$off2,off1 and $\Delta$off8,off1) are shown below off1. Corresponding differences are shown below on1. Note that eggs laid over the UV LEDs cause large differences when the LEDs are on (green arrows), including strong shadows on the chamber sidewalls (blue arrows).

**Supplementary Figure 3** Assessing tracking accuracy using the flies' trajectories and our suspicious jump detector.

**(a-b)** Detecting tracking errors. Our shadow detector (**Fig. 1h**) picks the detected ellipse closest to the center of each chamber; if the picked ellipse is wrong, a jump will appear in the trajectory. In fact, most tracking errors appear as such jumps. We hence examined the trajectories for suspicious jumps after each tracking and implemented a suspicious jump detector. (Tracking errors can also result in an increase in the number of flies – e.g., due to trajectories' being broken into pieces – or in flies' identities being swapped. When using Ctrax with extensions for our tracking tasks, both of these types of errors were rare.)  **(a)** Trajectories from tracking *without* background recalculation contain tracking errors.  The blue and green arrows point to multiple jumps (straight lines) to or from the center of the chamber (yellow arrow), with the blue arrow pointing to multiple almost identical jumps.  The fly did not jump to the center in these cases, however; instead, a change in grape juice level in the center well was mistaken for a fly (**Supplementary Fig. 2a**).  **(b)** Sample jump reported by our suspicious jump detector.  The image shows three consecutive partial (left chamber only) frames (i, i+1, i+2), with the ellipses showing where Ctrax reports the fly is and the arrow pointing to the fly's actual position in the error frame (i+1).  (Our suspicious jump detector reports two types of jumps of at least a certain length: a jump that is followed by another jump in approximately opposite direction and a jump with virtually no movement in the 30s before or after the jump, both uncommon and hence "suspicious" fly behaviors.)

**(c-d)** Auto-fix feature of our suspicious jump detector.  Our detector can fix a jump that coincides with an on/off state change and is followed by another jump to about the original position.  **(c-d)** Jump fixed by the detector, with the images showing three consecutive partial (left chamber only) frames (j, j+1, j+2) both before **(c)** and after **(d)** the fix.  The ellipses show where Ctrax reports the fly is and the white arrow points to the actual position of the fly in the error frame.  Strictly speaking, not Ctrax but the suspicious jump detector reported the positions in (**d**); the detector is MATLAB code separate from Ctrax.  Note that the LED in frame j+1 is not fully on (compare, e.g., the area of the LED pointed to by the blue arrows in frames j+1 and j+2), which caused the suspicious jump.

**(e)** Typical 1-hour trajectories of two (not-too-active) flies using Ctrax with extensions on a video with strong shadows.  Unlike in (**a**), there are no suspicious jumps to the same position.  There are some shorter incorrect jumps (arrows, easiest to see for blue arrow), usually caused by Ctrax's merging fly and shadow into a single object.  Our behavioral analysis based on the trajectories was essentially unaffected by these "minor" jumps, allowing us to ignore them.  Our "strong red light on/off" experiments did not have shadow problems.

# Figure 1

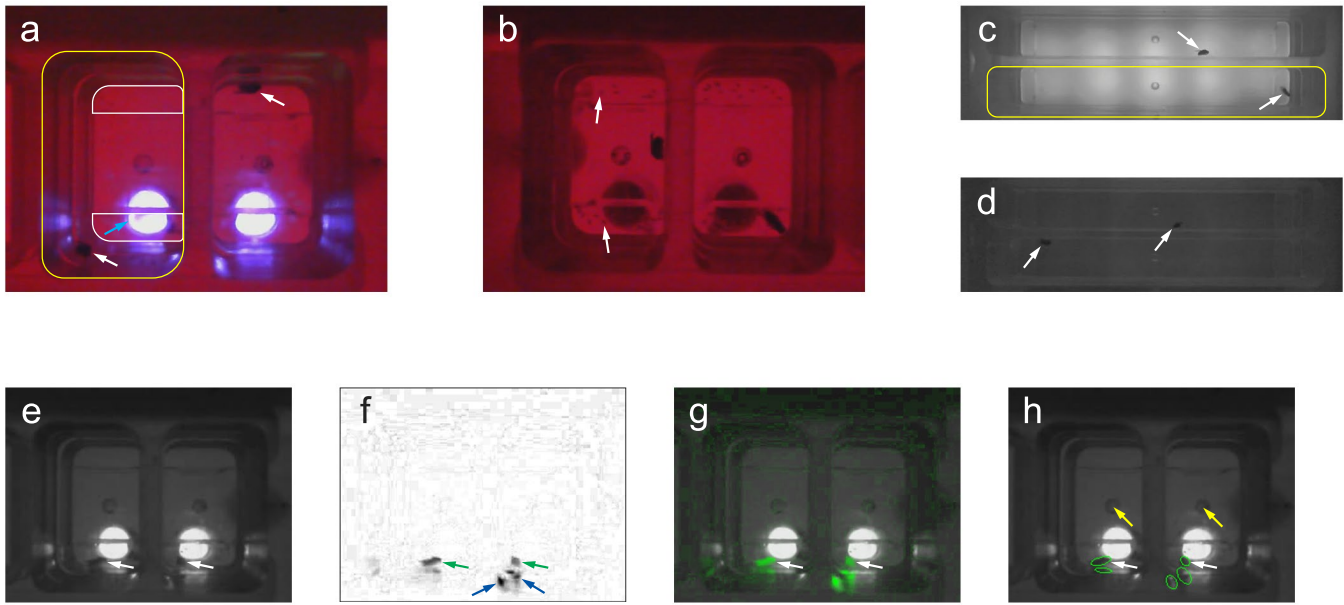

| sample 8h videos (from UV experiments) | | number of flies detected | | number of suspicious jumps (Ctrax with extensions) | |
|---|---|---|---|---|---|
| video number | UV | unmodified Ctrax | Ctrax with extensions | reported | tracking errors |
| 1 | on | 216 | 2 | 3 | 0 |
| 2 | on | 285 | 2 | 5 | 0 |
| 3 | on | 183 | 2 | 1 | 0 |
| 4 | on | 262 | 2 | 3 | 0 |

| sample 8h videos (from UV experiments) | | number of flies detected | number of suspicious jumps | | | |
|---|---|---|---|---|---|---|
| video number | UV | | auto-fixed | incorrect fix | reported (unfixed) | tracking errors |
| 5 | 1 min on/off | 2 | 26 | 0 | 2 | 0 |
| 6 | 2 min on/off | 2 | 14 | 1 | 1 | 0 |
| 7 | 5 min on/off | 2 | 4 | 0 | 6 | 0 |

# Supplementary Figure 1

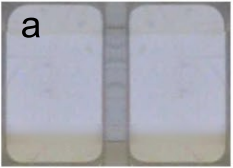 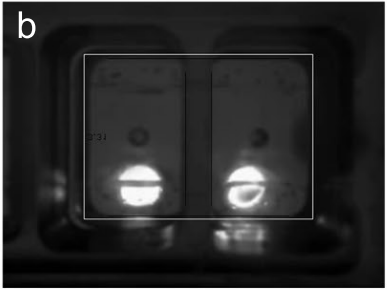

# Supplementary Figure 2

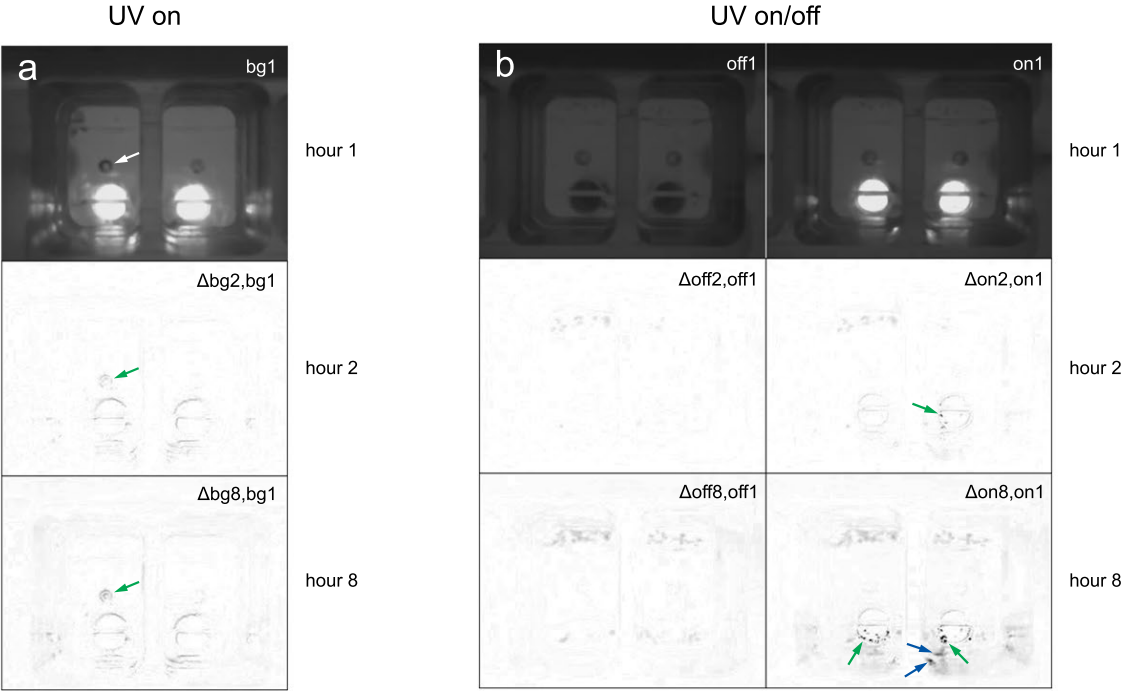

# Supplementary Figure 3

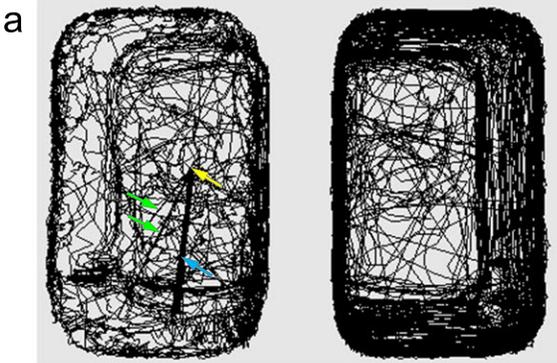

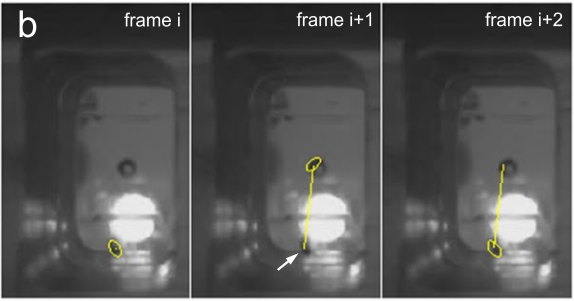

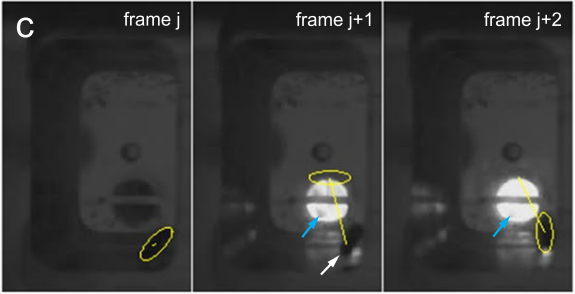

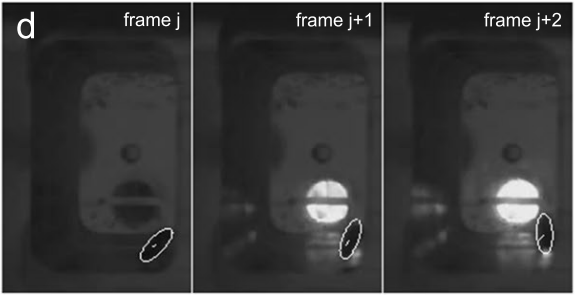

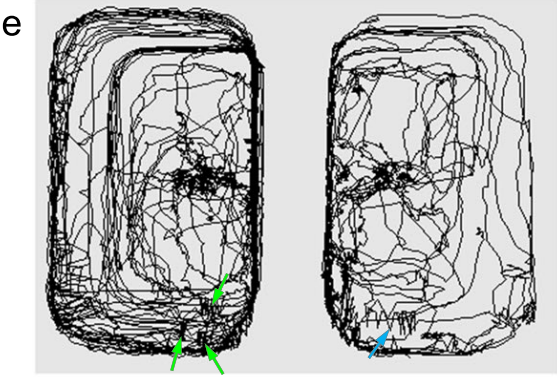